\documentclass[aps,prl,twocolumn,amsmath,amssymb,superscriptaddress]{revtex4-1}
\usepackage{graphicx}
\usepackage{amssymb}
\usepackage{natbib}
\usepackage{xfrac}
\usepackage{epstopdf}
\usepackage{float}
\usepackage{xspace}
\usepackage{xcolor}
\usepackage{spverbatim}
\usepackage{gensymb}
\usepackage{braket}
\usepackage{placeins}
\usepackage[normalem]{ulem}

\newcommand{\smthree}{Sm$^{3+}$\xspace}
\newcommand{\smtwo}{Sm$^{2+}$\xspace}

\begin{document}

\title{Orbital differentiation in a rare earth ion near half filling}

\author{D.~W.~Tam}
\email{tamd@ill.fr}
\affiliation{Institut Laue-Langevin, 38042 Grenoble Cedex 9, France}
\affiliation{Department of Applied Physics, KTH Royal Institute of Technology, SE-106 91 Stockholm, Sweden}
\author{\O.~S.~Fjellv\aa g}
\email{oystein.fjellvag@ife.no}
\affiliation{Institute for Energy Technology, Department for Hydrogen Technology, 2007 Kjeller, Norway}
\author{B.~Gonano}
\affiliation{Chemistry Department and Center for Material Science and Nanotechnology, University of Oslo, Oslo NO-0315, Norway}
\author{E.~X.~M.~Trewick}
\affiliation{School of Chemical and Physical Sciences, Victoria University of Wellington, P.O. Box 600, Wellington 6140, New Zealand}
\affiliation{MacDiarmid Institute for Advanced Materials and Nanotechnology, P.O. Box 600, Wellington 6140, New Zealand}
\author{M.~Månsson}
\affiliation{Department of Applied Physics, KTH Royal Institute of Technology, SE-106 91 Stockholm, Sweden}

\date{\today}

\begin{abstract}

In the electronic structure of rare earth ions, details can be crucial for understanding the origin of correlated states.
Currently, state-of-the-art wavefunction-based methods (e.g. CASSCF) can capture single-ion properties of molecular magnets and some band insulating systems, but are expensive to perform, and cannot be applied to metallic systems.
Here, we carry out full multiplet calculations for SmN with the cubic rock-salt structure using the crystal field software Quanty.
The calculated experimental observables from our model agree precisely with magnetic susceptibility and inelastic neutron scattering experiments, and the spatially resolved spin density also agrees with DFT predictions.
By tuning the crystal field parameters, our full multiplet-derived wavefunctions reveal the presence of orbital differentiation of the ground $J=\sfrac{5}{2}$ manifold into two groups, which are distinct subspaces with different orbital stiffness.
This violation of Hund's rule, observed with a fast and inexpensive calculation tool, represents a step forward for microscopic models of rare earth materials, and may provide support for the claim of unconventional triplet superconductivity in SmN.

\end{abstract}

\maketitle

\subsection{Introduction}

In rare earth materials, a microscopic description of the single-ion properties usually starts from the large Coulomb repulsion $U$, which is the mutual electrostatic repulsion among electrons in the $f$ shell of one ion \cite{Joha79,HoUe03,PDTK14}.
Applying Hund's rules, a manifold of total angular momentum $J$ can be identified as the ground state, which is further split at lower energy scales by spin-orbit coupling (SOC) and the crystal electric field (CEF).
It is common in the literature to find microscopic descriptions of the $f$ subsystem within a degenerate subspace where, very often, $J$ is taken to be fixed and of integer or half-integer value.
However, the importance of explicitly considering $U$, SOC, and CEF simultaneously in rare earth materials, the so-called full multiplet approach, has continued to gain traction in the modeling community, especially as the desire for accuracy in high-throughput methods has increased \cite{TVFS85,LNFM25}.

The situation for \smthree ($4f^5$), near half filling, is among the most sensitive to the detailed balance between the different interactions.
The ground manifold of \smthree from Hund's rules is a Kramers doublet with $S=\sfrac{5}{2}$, $L=5$, $J=\sfrac{5}{2}$, while the first excited multiplet with $J=\sfrac{7}{2}$ is well-known to exhibit a small separation of only $\Delta \approx 150$ meV \cite{DiCr63,MuHa22}, which is the smallest of any trivalent rare earth, except Eu where $J=0$.
This implies that electrostatic interactions induce substantial inter‑multiplet mixing in \smthree, and therefore that a full‑multiplet treatment is required to capture both the ground state wavefunction and its response to the crystal field.
Crystal field softwares that do not explicitly include these Coulomb interactions, involving diagonalization of the $\binom{14}{5}=2002 \times 2002$ matrix Hamiltonian, are not expected to work well for \smthree.
Experimentally, it has already been recognized that the full-multiplet requirement applies to essentially every material with \smthree \cite{SHLK15,SYCK18,PFSA19,TCKP23}, making it more conceptually similar to the actinides and Pr${}^{4+}$ \cite{DRKC21,RKSB23,RWML23} than to other rare earths.
In those actinide systems, intermediate valence character in the ground state often involves delocalization of an integer number of electrons \cite{PSTS02,TrGP12,DDSJ13}.
On the other hand, $4f$ materials with intermediate valence are found in fractional valence states \cite{BDCH11,BMJB14,KGKZ18}, as is the case with a large percentage of Sm-based materials \cite{SHLK15,HYMA18}.
It is therefore interesting to consider cases involving \smthree where CEF is large and competing with SOC, where full-multiplet modeling might provide information on how the partial delocalization process can occur.
In semiconducting SmN \cite{ASHT04,DSMD07,NRPT13} (nominal ligand charge $-3$), the crystal field is expected to be among the strongest available.

In some cases, highly accurate models of rare earth magnets have been successfully obtained using the complete-active-space self-consistent-field (CASSCF) method, which is an \textit{ab initio} wavefunction method \cite{BRHS20,BWPB21,RKBB23,BhBH26}.
These successes include the identification of an orbital differentiation mechanism between bonding and anti-bonding Nd $f$ orbitals in a single-ion magnet \cite{WSYJ16,Fchi25,JGFJ25}.
The use of CASSCF therefore represents a significant improvement over density functional theory (DFT), which typically makes crude assumptions for static electron correlations based on a single reference point.
%Indeed, DFT-based approaches applied to rare earth materials often fail to capture anything beyond an approximation of the local electronic density.
CASSCF overcomes these limitations by applying a multi-reference approach within the space of active electronic orbitals, and can therefore succeed in capturing detailed properties.
However, CASSCF remains only infrequently applied to bulk magnetic materials \cite{DBMP17}.
One reason is that CASSCF methods are very computationally expensive.
There is also the conceptual difficulty that the active space Hamiltonian is applied \emph{before} considering spin-orbit coupling \cite{VaCS16}, which can make interpretation difficult. %(Kohn, Heuvel)
One of the most important limitations remains that CASSCF is not practical for metallic materials, and possibly semiconductors, since the active space cannot be constructed locally from hybridized bands with extended states.
Finally, the crystal-field environment is not an independently tunable part of the CASSCF calculation: changes to the crystal field are entangled with the self-consistent electronic structure, orbital optimization, and embedding assumptions.
Therefore, measuring the orbital response to perturbations of the local potential, giving a measure of their stiffness in dynamical processes, remains out of reach for such an approach.

Here, we take an alternative approach using the full-multiplet crystal field software \verb+Quanty+ \cite{Have16}.
The advantage of crystal field software is that the active space is controlled by angular-dependent potentials operating within a fixed radial wavefunction, which eliminates much of the configurational complexity associated with the radial degree of freedom.
The active space defined in \verb+Quanty+ is a set of $N-$particle configurations, and the potentials are applied using matrix methods, including two-body Coulomb interactions, but without defining a local density.
Therefore, the use of \verb+Quanty+ greatly simplifies the computational expense of the calculations.
It can also be applied to metallic materials, which makes it a significantly more versatile tool compared to the \textit{ab initio} wavefunction methods.
Most conveniently, it allows the crystal field potential to be directly parameterized as a set of spherical harmonics with tunable coefficients, rather than solving a complete set of self-consistent calculations.

In this work, we show that the full-multiplet approach in \verb+Quanty+ can efficiently capture the presence of orbital differentiation in SmN.
For rare earth materials, the small radius of the magnetic $4f$ shell is expected to provide very accurate estimates of the radial expectation values of the real wavefunctions, as is evidenced by the plethora of examples in the literature of crystal field models constructed this way.
Our identification of orbital differentiation in SmN is therefore likely to play a significant role in understanding the magnetic ground state in the real material.
Our method shows a new way that \verb+Quanty+ can go beyond modeling energy levels and core spectroscopy for rare earth materials, providing a rapid diagnostic tool for evaluating when the magnetic subspace is no longer the perfectly degenerate subspace of Hund's rules.

\subsection{Modeling details}

To model SmN, we consider the full-multiplet single-ion Hamiltonian

\begin{align}
H &= H_U + H_\text{SOC} + H_\text{CEF}  \notag  \\ 
&=  H_U + \zeta \mathbf{L} \cdot \mathbf{S} + \sum_{l\in\{2,4,6\}}  \sum_{m=-l}^l A_l^m C_l^m(\theta,\phi)
\label{ham}
\end{align}

where $A_l^m$ are (Wybourne) coefficients of the tesseral harmonics $C_l^m(\theta,\phi)$ with even parity.
Physically, the angular part of the electrostatic potential acts on the wavefunction through two other factors that can be explicitly included by defining $B_l^m = A_l^m \theta_l \langle r^l \rangle$, where $\theta_l$ is the Stevens factor that projects the potential on a shell of total angular momentum $J$, and the radial expectation values $\langle r^l \rangle$ are usually tabulated from Hartree-Fock calculations \cite{Stev52,FrWa62}.
However, treating the problem directly with $A_l^m$ avoids making a choice of $\langle r^l \rangle$, making it a more straightforward choice for modeling.

The crystal field in a cubic environment is almost trivial, containing only two independent coefficients $A_4^0$ and $A_6^0$, with other $m$ terms at given $l$ given by exact relations. % $A_4^4=\sqrt{5/14} A_4^0$, $A_6^4=-\sqrt{7/2} A_6^0$.
In \smthree, this potential splits the $J=\sfrac{5}{2}$ ground manifold into a doublet and quartet.
At the same time, the effect of $A_6^m$ takes a special meaning in \smthree, because a potential of rank~6 cannot act on a manifold with $J=\sfrac{5}{2}$ \cite{Stev52,Hutc64} and therefore $A_6^0$ enters only through the inter-multiplet mixing with the first excited $J=\sfrac{7}{2}$ multiplet.
Therefore, by virtue of this simple CEF environment with a straightforward interpretation, full-multiplet modeling efforts are highly likely to succeed at capturing the effects of interplay between the $U$, SOC, and CEF degrees of freedom in this multi-$f$-electron ion.

\begin{figure*}[tbph]
\includegraphics[width=.85\textwidth]{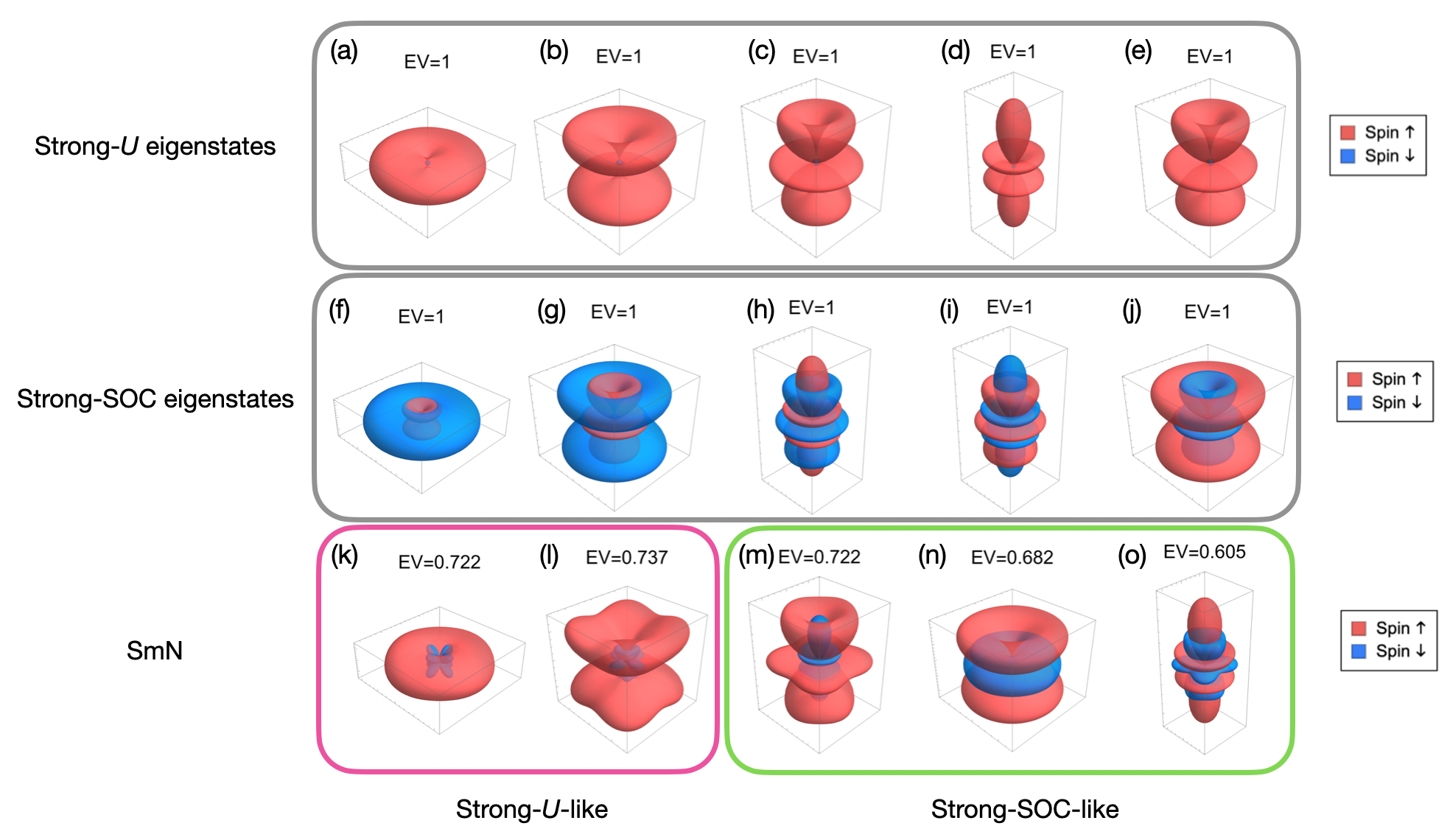}
\caption{
Magnitude $|\psi|$ of the one-electron orbital wavefunctions (eigenvectors of the density matrix projected on our quantization axis) for three models of \smthree.
(a-e) Eigenvectors of the model containing only the Coulomb interaction with $U=80$~\% of the Cowan code values, with the eigenvalues (EV) shown above whose physical meaning is an occupation factor. The spin is represented by the surface color of red or blue.
(f-j) Eigenvectors of the model containing only a spin-orbit coupling (SOC) term with $\zeta=75$~meV (about half the real value).
(k-o) Eigenvectors of the complete model of SmN, containing the Coulomb $U$, spin-orbit coupling ($\zeta=150$~meV) and the crystal field potential ($A_4^0=200$~meV, $A_6^0=20$~meV).
Visually, it is already apparent that two of the orbitals (k-l) resemble the $U$ eigenstates, while the other three orbitals resemble the SOC eigenstates.
}
\label{figlimiting}
\end{figure*}

We start by considering the effects of limiting cases (strong $U$ or strong SOC only) before proceeding to SmN.
In all cases, we diagonalize Eq.~\ref{ham} in \verb+Quanty+ \cite{Have16} to obtain the ground state wavefunction within a density matrix description.
We then further analyze the multi-$f$-electron ground state with 5 electrons by diagonalizing the density matrix, which leads to 5 large eigenvalues.
The associated eigenvectors give the natural one-electron orbitals, and the eigenvalues are their occupation factors.
Here, the diagonalization is not used as a calculational convenience, but as a diagnostic that reveals a physically meaningful partitioning of the $4f^5$ shell into orbital subsets.
By tuning the CEF parameters $A_4^0$ and $A_6^0$ over a wide range, we discover that two of the natural orbitals of SmN remain nearly the same as the eigenstates of $H_U$, unchanged by the different crystal fields over wide regions of parameter space.
The other three natural orbitals are highly similar to eigenstates of $H_\text{SOC}$ and are continuously tunable with $A_4^0$ and $A_6^0$.
These results indicate that the ground state wavefunction exhibits internal structure that divides the 5 constituent electrons into two groups, and can no longer be considered a perfectly degenerate subspace.
The different orbital stiffness for the two groups may have relevance for the behavior of other Sm-based systems, which are often enigmatic \cite{SKVS05,HYMA18}.
Since SmN in particular has been argued to be a ferromagnetic superconductor in thin films \cite{AGEC16,HBRN19}, we focus our discussion on the relationship of orbital differentiation to the delocalization mechanism in metallic hosts.

\subsection{Results for limiting cases}

Our models are constructed including $H_\text{CEF}$, $H_\text{SOC}$ with $\zeta \approx 150$ meV, and $H_U$ with the coefficients reduced to 80\% of the values produced via the Cowan code \cite{Rete24}, which is a typical practice \cite{SYCK18,SHAA16}.
For SmN, we construct an \textit{ab initio} model for $H_\text{CEF}$ using the \verb+PyCrystalField+ software \cite{Sche21} to calculate the initial Stevens $B$ operators using the point charge construction, which we convert to the tesseral harmonic expansion coefficients $A$ for direct input in \verb+Quanty+.
We use (001) as the quantization axis, and our orbital shapes show the projection of the states along this axis.
All calculations are carried out in a small external magnetic field of +10~Oe oriented along the (001) direction, simply to resolve the quantum numbers separately for the two states in the Kramers doublet.

In Fig.~\ref{figlimiting}(a-j), we show how our approach handles limiting cases.
In the limit of no crystal field and no SOC, $H_U$ favors the ferromagnetic alignment of spins (Hund's coupling).
In this case, each single electron occupies an $f$ orbital wavefunction of definite $M_L$ that is simply one of the spherical harmonics $Y_3^m$, as shown in Fig.~\ref{figlimiting}(a-e).
In the other limit, where SOC is dominant and $H_U$ is ignored, SOC mixes the $M_L$ states to form good eigenstates of $J$, shown in Fig.~\ref{figlimiting}(f-j).
These states couple together neighboring $M_L$ states with opposite $M_S$ (thereby forming the same $M_J$), in a combination given by the Clebsch-Gordan coefficients (for tables, see Supplementary Material \cite{SM}).
Overall, our natural orbitals for the limiting cases of strong $U$ or strong SOC are highly consistent with expectations.

To further validate our full-multiplet model, we carried out electronic structure calculations with spin-polarized density functional theory (DFT) using the PBE generalized gradient approximation and Hubbard correction (PBE+U), similar to other calculations previously used to describe the electronic and structural properties of many of the rare earth nitrides \cite{LLCS07,Holm21,HTTB24,MVKH25}.
The computational challenges associated with capturing the strongly localized $f$ electrons is a well-known issue in the PBE+U approach.
However, since the effects of the electrostatic crystal field potential, $H_\text{CEF}$, are present in both PBE+U and our full-multplet \verb+Quanty+ models, a comparison of the effects of this potential serves as another important validation to the full-multiplet approach.
Our DFT calculations are carried out with Quantum ESPRESSO \cite{GBBC09,GABB17} (for details, see Supplementary Material \cite{SM}), and in Fig.~\ref{figdft}(a), we show the $z$ projection of the total spin density, $M_S^{(z)}(\mathbf{r})$, throughout one SmN unit cell.
We further evaluate the same spin density $M_S^{(z)}(\mathbf{r})$ in Fig.~\ref{figdft}(b), which is calculated directly from the density matrix for one Sm ion in the basis of $f$ orbital states.
For the corresponding \verb+Quanty+ equivalent to the PBE+U Hamiltonian, we set $H_U=H_\text{SOC}=0$, and also force all $f$ electrons to be of the same spin.
The density matrix projection $M_S^{(z)}(\mathbf{r})$ for this model is shown in Fig.~\ref{figdft}(c), which bears a strong similarity to the DFT solution in Fig.~\ref{figdft}(b) and shows that solutions obtained by both methods are equivalent.
Overall, this equivalence confirms that it is reasonable to model the $f$ electron shell as localized and interacting with the expected $H_\text{CEF}$ potential, and also emphasizes that the inclusion of SOC and 2-electron Coulomb interaction operators in the full Hamiltonian within \verb+Quanty+ represents a substantial improvement in the treatment of the $f$ electrons over that in PBE+U.

\begin{figure}[tbph]
\includegraphics[width=.47\textwidth]{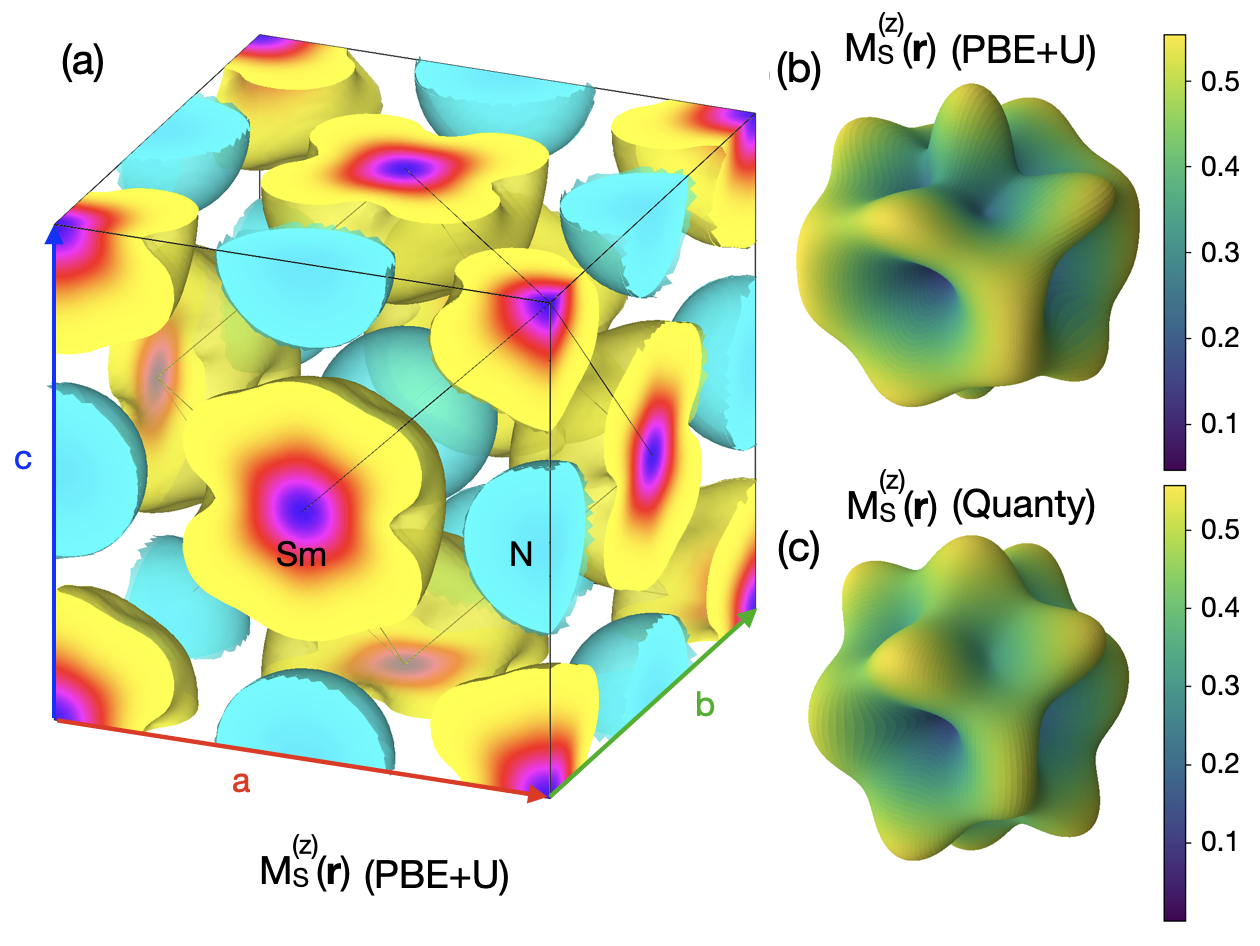}
\caption{
Equivalence of DFT and Quanty approaches for the bare $H_\text{CEF}$ interaction.
(a) z component of spin density $M_S^{(z)}(\mathbf{r})$ from DFT calculations on the real space density grid, and (b) $M_S^{(z)}(\mathbf{r})$ reconstructed from the DFT density matrix projected onto the orbital basis.
(c) $M_S^{(z)}(\mathbf{r})$ constructed from our restricted Quanty model density matrix projected onto the orbital basis.
}
\label{figdft}
\end{figure}

\begin{figure*}[tbph]
\includegraphics[width=.97\textwidth]{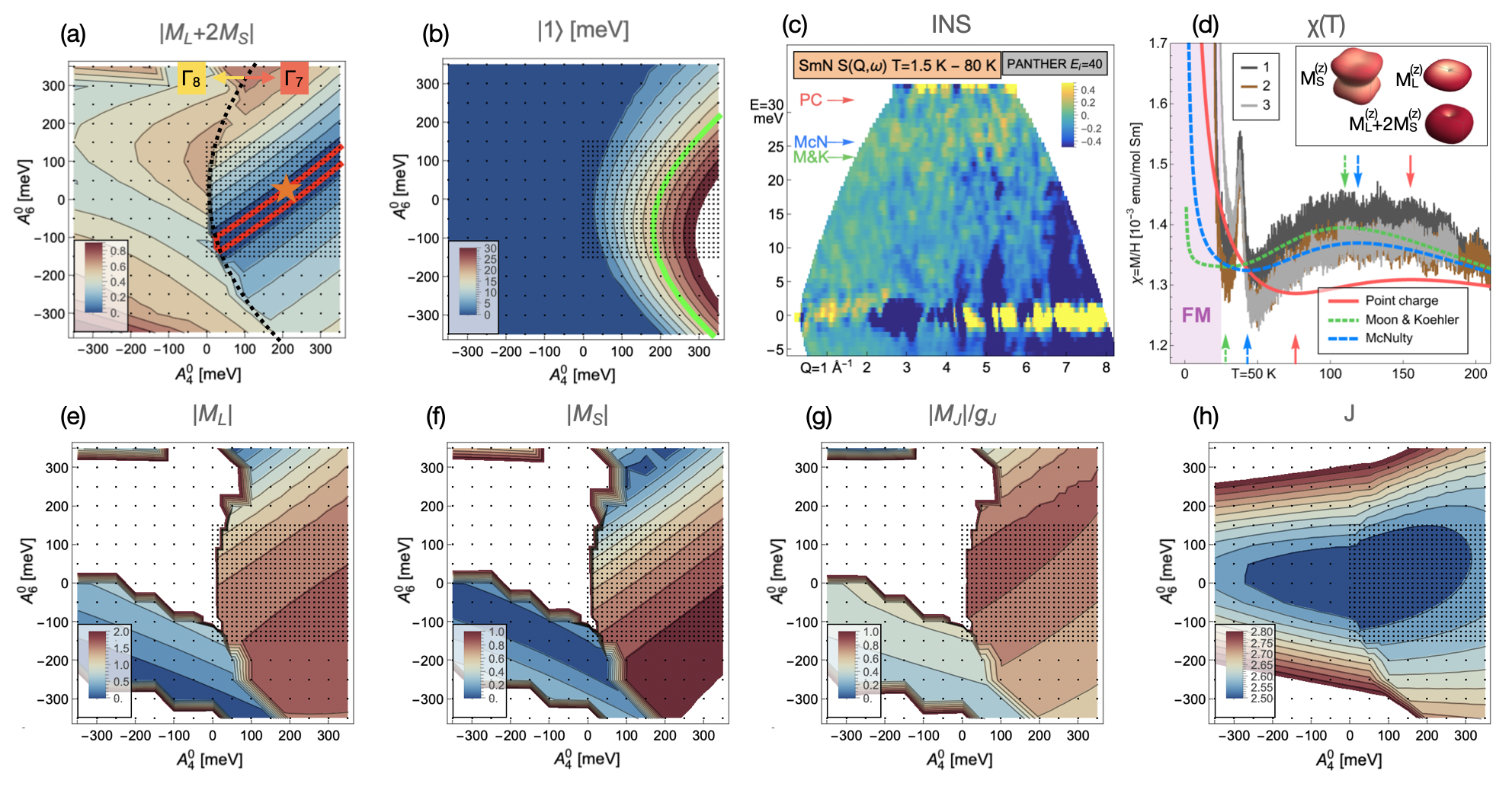}
\caption{
Calculated properties for SmN and their comparison to experiments. In (a,b,e,f,g,h), black points represent separate calculations that were interpolated to create the color plot, and white regions contain values outside the color scale given in the inset.
(a) Magnetic moment $|M_L+2M_S|$. The black dashed line indicates the regions with the $\Gamma_8$ quartet ground state or $\Gamma_7$ doublet ground state. The red line represents the region with vanishing magnetic moment (in experiments, $M=0.035$~$\mu_B$).
(b) Energy splitting between the crystal field ground doublet $\ket{0}$ and first excited doublet $\ket{1}$. The green line represents the experimental value reported in Ref.~\cite{MoKo79}.
(c) Temperature-subtracted inelastic neutron scattering (INS) measurements at Panther with a flat plate (reflection geometry) using $E_i=40$~meV. The expected splitting from the three models are labeled $PC$ (point charge), $M\&K$ (Moon \& Koehler \cite{MoKo79}), and $McN$ (McNulty \cite{McRT16}).
(d) Three independent measurements of the magnetic susceptibility $\chi(T)$ were carried out on a 64.6~mg powder sample in a field of 100~Oe. The three model predictions are also shown, with arrows indicating the calculated maxima and minima. The purple region represents FM order. Inset: the calculated shapes of the spin $M_S^{(z)}$, orbital $M_L^{(z)}$, and anisotropy of the total magnetization density $|M_L^{(z)}|+2|M_S^{(z)}|$ of the \smthree ion projected along the $z$ quantization axis.
(e-h) Evolution of $|M_L|$, $|M_S|$, $|M_J|$, and $J$ remain smooth over the $\Gamma_7$ region.
}
\label{figsmn}
\end{figure*}

\subsection{Results for $\boldsymbol{\mathrm{SmN}}$}

SmN is ferromagnetic below $T_C \approx 25$~K and is known to exhibit an extremely small ground state magnetic moment ($M=0.035$~$\mu_B$), which is orbital-dominated in the sense that the magnetic orbital moment $M_L$ aligns to an external field \cite{MRZG08,AHMN21,MTAV25}.
The small moment has also been confirmed by measurements of the magnetic form factor in neutron scattering \cite{MoKo79}.
In thin films, a low-resistance state has also been observed at lower temperatures, which could be superconductivity \cite{AGEC16,HBRN19}.
As it is not known whether this low-resistance state is connected to the origin of the small moment, a precise description of the orbital wavefunctions and magnetic anisotropy is highly desired.
Recently, it was determined that the valence of Sm remains Sm${}^{3+}$ in the bulk of SmN at all temperatures, whereas the surface forms Sm${}^{2+}$, confirming that an intermediate valence configuration is also nearby in energy. \cite{KCTC26}.

Two proposals for crystal field models of SmN were put forward in Ref.~\cite{MoKo79} (Moon \& Koehler) and Ref.~\cite{McRT16} (McNulty), both finding parameters $A_4^0 \approx 200$~meV and $A_6^0 \approx 10$~meV.
Both models attempted to introduce an additional magnetic exchange parameter to explain magnetic susceptibility measurements.
However, in the case of Ref.~\cite{MoKo79} it was not known that the system forms a ferromagnetic ground state below $T_C \approx 25$~K, leading to incorrect assumptions about the strong upturn in susceptibility there.
In the case of Ref.~\cite{McRT16}, the exchange was estimated from Curie-Weiss fits over a large temperature range, which is problematic due to the level splitting of about $200$~K between the doublet and quartet states of the ground manifold that generates non-monotonic behavior in the susceptibility.

Using the parameters from either model, but treated with the full-multiplet approach (Eq.~\ref{ham}) and without any exchange interactions, we find good agreement between all three models and available experimental results, with the best agreement for the McNulty model (Ref.~\cite{McRT16}).
We show the resulting natural orbitals from our model in Fig.~\ref{figlimiting}(k-o), which are mixed between the Coulomb and SOC orbitals from the limiting case models.
The largest occupation factor in our model for SmN, EV=0.737, corresponds to the one-electron orbital with the largest rotational symmetry breaking imposed by the crystal field (Fig.~\ref{figlimiting}[l]).
This is intuitively reasonable, because this electron orbital shape closely matches the shape of the crystal field potential, giving it the highest stability.

In Fig.~\ref{figsmn}, we show in detail how our model for SmN depends on the crystal field parameters $A_4^0$ and $A_6^0$ and compare it to experiments.
The red contour line in Fig.~\ref{figsmn}(a) indicates a boundary region where the magnetic moment $M_L+2M_S$ of the $\Gamma_7$ doublet crosses through zero, indicating a switch from orbital-dominated to spin-dominated.
The energy of the doublet-quartet excitation was reported in Ref.~\cite{MoKo79} to be near 19~meV, which we show as a green contour line in Fig.~\ref{figsmn}(b).
Therefore, the crystal field parameters must lie close to the intersection of these lines, near the orange star in Fig.~\ref{figsmn}(a).

To verify these parameters independently, we synthesized a powder sample of SmN and carried out inelastic neutron scattering (Fig.~\ref{figsmn}[c]) and magnetic susceptibility measurements (Fig.~\ref{figsmn}[d]).
We find evidence for a broad level near $E=26.2 \pm 0.3$~meV  with the unique $Q$-dependent magnetic form factor of \smthree \cite{MoKo79,Love86,SM}, which is slightly different from the 19~meV reported in Ref.~\cite{MoKo79}.
The full-multiplet nature of the calculations also captures the broad ``hump''-like maximum in the susceptibility that peaks near $T=125$~K and has a minimum near $T=50$~K before the onset of FM order.
The peak near $T\approx125$~K represents the change in magnetic susceptibility as the system crosses from an effective $J=\sfrac{5}{2}$  model (representing the full ion) into the $\Gamma_7$ magnetic doublet ground state, similar to the results of SmCoIn$_5$, SmI$_3$, and other Sm-based compounds with the doublet ground state \cite{IKYN22,TCKP23,HYMA18}.
Our susceptibility data also display a sharp hump in the data at a temperature near $T=40$~K, whose origin remains unknown, but is consistent with a very small amount of oxygen inside the sample environment, or an incipient-like order related to the FM order below $T_C=25$~K.
Nevertheless, it is too sharply peaked in temperature to be related to any crystal field effects.
However, aside from this small hump, the ``McNulty'' model of Ref.~\cite{McRT16} with our full-multiplet treatment provides the best agreement with the data in the paramagnetic phase, including the broad minimum at $T=50$~K and the broad maximum at $T=125$~K.
Overall, using this model with the full-multiplet treatment, but without additional exchange interactions, closely captures all of our new experimental data, as well as other physical observables reported previously (see Supplementary Material \cite{SM}).
Therefore, we conclude that our full-multiplet model of SmN is highly accurate.

\begin{figure*}[tbph]
\includegraphics[width=.97\textwidth]{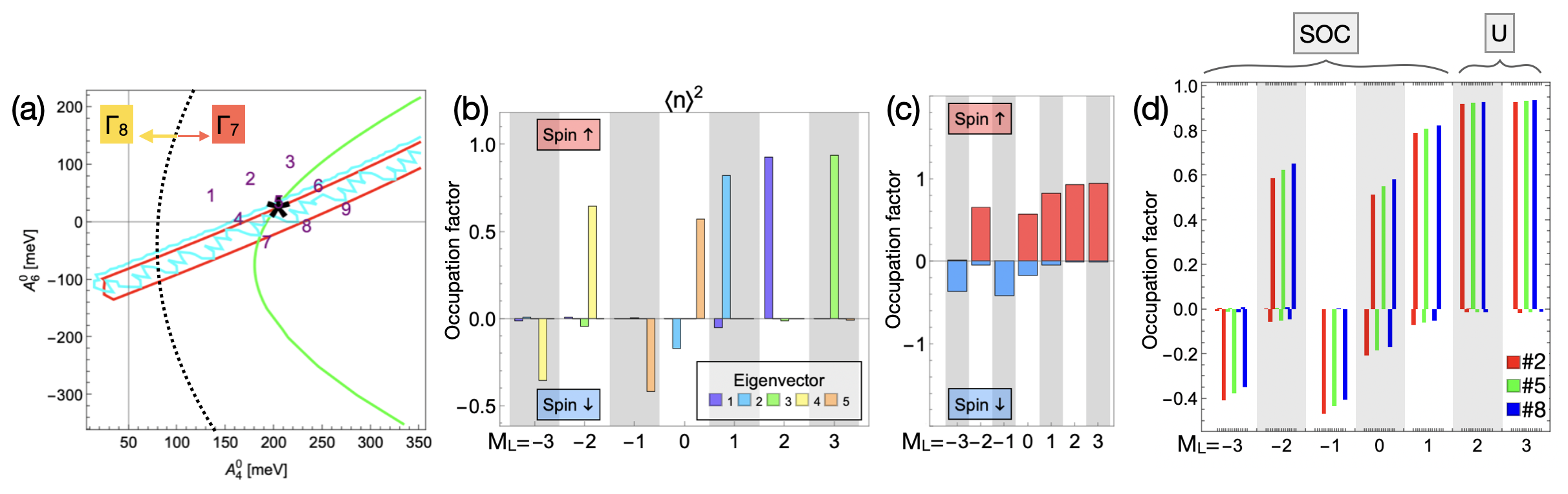}
\caption{
Orbital content of SmN.
(a) Overall phase diagram containing the annotations from Fig.~\ref{figsmn}(a-b) and the region of vanishing magnetic moment shown in light blue (for additional annotations, see Supplementary Material \cite{SM}).
The positions of nine individual calculations (\#1-\#9) are indicated with numbers.
(b) Projection of the natural orbitals from calculation \#5 into the basis of orbital states with definite $M_L$ (spherical harmonics), with values given by occupation number. Positive (negative) values indicate a spin up (spin down) component.
(c) Total occupation of each $M_L$ basis state combining all natural orbitals from calculation \#5.
(d) Total occupation of each $M_L$ basis state, similar to panel (c), for three calculations (\#2, \#5, \#8) whose parameters are shown in (a). The ``Coulomb-like'' natural orbitals and ``spin-orbit-like'' natural orbitals are easily distinguished because they occupy distinct $M_L$ states in SmN.
}
\label{figorbdiff}
\end{figure*}

\subsection{Evidence for orbital differentiation}

We now proceed to evaluate how the one-electron ``natural'' orbitals in SmN evolve in the parameter space of $A_4^0$ and $A_6^0$.
In Fig.~\ref{figorbdiff}(a), we show the position of nine individual calculations labeled \#1-\#9, with contour lines from Fig.~\ref{figsmn} overlaid to clarify the position of these calculations.
The central calculation, \#5, is the model matching the experiments and whose natural orbitals and occupations (eigenvectors of the five leading eigenvalues) are depicted in Fig.~\ref{figlimiting}(k-o).
The other calculations are chosen along lines parallel to the boundary where $M_L+2M_S=0$.
In all calculations, we find that the 6th-leading state has an occupation significantly smaller than the first five, while the spin density associated with it is very small, which is also the case for the combination of states 7-14 (for details, see Supplementary Material \cite{SM}).
We therefore focus entirely on the five leading states with large occupation.

Using the states of definite $M_L$ as a basis (the spherical harmonics), we project each of the five natural orbitals into this basis as shown in Fig.~\ref{figorbdiff}(b), where the spin content is represented by placing the bar above or below the axis for spin-up or spin-down, respectively.
Two of the natural orbitals, states 1 and 3, are completely and uniquely placed in states of definite $M_L$, which are the states with large $M_L=+2$ and $+3$ as expected from Hund's rule.
On the other hand, the other three natural orbitals are mixtures of neighboring $M_L$ states with opposite spin-up and spin-down components, closely resembling the SOC eigenstates.
Because each natural orbital is associated with a distinct $M_L$ state, resulting in no overlapping bars in Fig.~\ref{figorbdiff}(b), the total wavefunction can be represented more compactly by combining the five natural orbitals together, which is shown in Fig.~\ref{figorbdiff}(c-d).

Evaluating the other calculations \#1-\#9 with different crystal fields, we find essentially no differences in the composition of the natural orbitals when moving in a direction parallel to the green contour line representing constant energy splitting, for instance between calculations \#4, \#5, and \#6 (for all calculations, see Supplementary Material \cite{SM}).
However, we do find differences when moving along the perpendicular direction, for instance comparing calculations \#2, \#5, and \#8, which is shown in Fig.~\ref{figorbdiff}(d).
In this case, the two Coulomb-like states remain unchanged by the crystal field, which leads to the same values in Fig.~\ref{figorbdiff}(d) for all three calculations.
On the other hand, the three SOC-like states are continuously tunable, with the occupation of each $M_L$ state changing by around 0.05 in occupation number in this case (left axis of Fig.~\ref{figorbdiff}[d]).
As the doublet-quartet separation is increased through the action of the crystal field, the three tunable states move closer to the Hund's rule configuration of all spins aligned, consistent with a weaker role for SOC when the crystal field energy is larger.
However, no change is observed for the Coulomb-like states.
These observations allow us to identify a mechanism of orbital differentiation in SmN, wherein two orbitals remain locked to the their strong-$H_U$ limits, and the other three are allowed to vary with changes in the local electrostatic potential.

\subsection{Discussion}

Orbital differentiation occurs in SmN because of the crystal field.
It results in two natural orbitals with ``Coulomb-like'' states of definite $M_L$ (spherical harmonics), similar to the expectations of Hund's rule.
The other three natural orbitals are ``SOC-like'' mixtures of $M_L$ and $M_S$ states that resemble the good eigenstates of the total angular momentum $J$, and which are continuously tunable by changing the crystal field.
Our method successfully explains the properties of SmN in the paramagnetic state using the full-multiplet approach, highlighting the importance of treating the local Coulomb, spin-orbit, and crystal field potentials on equal footing for rare earth ions close to half filling.

Orbital differentiation at a single-ion level draws a comparison between SmN and multi-band $d$-electron systems, where it has been recognized that the Coulomb or Hund's coupling terms are often essential to capturing basic details about strongly correlated ground states \cite{HaKo09,GeMM13,MeGC14,MYWL16,LYWT16,SFMK17,YZSL17,TYXW20}.
A prominent example of this effect occurs in iron-based superconductors \cite{HaKo09,FCDF22}, where the Hund's coupling is responsible for a crossover from an incoherent regime at high temperature, to coherent and strongly correlated dynamics at low temperature.
At the same time, the role of spin-orbit coupling has been identified as a key ingredient in correlated ground states in some $d$ systems, such as iridates \cite{KJMK08,WCKB14}, and $4d^1/5d^1$ systems with Mo${}^{5+}$, Os${}^{7+}$, and Re${}^{6+}$ \cite{XBPF16,MTSM16,UAFN21,CMCW24}.
Extending the family of orbitally differentiated materials to multi-$f$-electron systems near half filling represents a way to join the regimes of strong Hund's coupling and strong SOC, and may lead to new types of tuning knobs in samples with \smthree.
In addition, closer proximity of the magnetic electrons to the nucleus in rare earths also better isolates the effects of electronic correlations from lattice degrees of freedom, such as the occupation effects expected from $d$-electrons in Jahn-Teller active environments \cite{StKh20}.

SmN itself may be an example of such a tuning knob at work.
A semiconductor at most temperatures, it has been previously argued to be a superconductor of heavy fermion type below $T_c \approx 4$~K in thin films.
If this is the case, it is interesting to consider what role the orbital differentiation may have in the context of electronic delocalization.
The relevance of orbital differentiation in this context is that it constrains which subset of $4f$ degrees of freedom can participate in delocalization or hybridization processes, rather than providing a complete microscopic theory of superconductivity.
As such, an examination of the role of an orbital differentiation mechanism is not at all firmly established, but merely provides a template for the orbital subspace that might be used to construct a microscopic theory.
Sm-based systems frequently exhibit a binding energy of $E_B \approx 6$~eV for the $4f^5$ configuration (\smthree) and $E_B \approx 0$ for the $4f^6$ configuration (\smtwo), which is nonmagnetic with $L=3$, $S=3$, and $J=L-S=0$ according to Hund's rules \cite{TCKP23,CGKG17,TCAS24}.
Due to the proximity of the $4f^6$ states with the Fermi energy, hybridization between the $f$ states and the conduction electrons can occur, which is a hole-like delocalization transition, involving the attachment of an electron to the $f$ site.
If such hybridization is magnetically-driven (Kondo coupling), such as in Ce- and Yb-based systems, the host ion is regarded as part of a magnetic singlet with the conduction states \cite{WhTM15}.
On the other hand, the hybridization process has been shown to be different in Sm-based systems, because the electronic specific heat coefficient $\gamma$ is often orders of magnitude larger than what is predicted from the number of delocalized states \cite{HYMA18}.
This appears to indicate that the hybridization process in Sm involves charge-like degrees of freedom, instead of or in addition to magnetic ones.

In our orbitally differentiated picture of Sm, the hybridization process should not involve the two Coulomb-like orbitals, because they are rigidly fixed by the crystal field with constant occupation under perturbations of the local electrostatic field.
Therefore, hybridization can only occur within the submanifold of the three SOC-like orbitals.
For a $4f^6$ configuration with $S=3$ to occur in SmN (the Hund's rule \smtwo configuration), one can ask what the wavefunction of the sixth electron would need to be, in order to minimize the energy cost of localizing it from the metallic host.
As shown in Fig.~\ref{figorbdiff}(b), the required orbital occupation of the sixth electron can never satisfy this condition, since it would need to cancel the existing orbital magnetic moment in the spin-down channel, while removing all the occupation of the $M_L=-3$ orbital.
In our model, the sixth natural orbital, which is unoccupied in the $4f^5$ \smthree wavefunction, is an SOC-like orbital that mainly occupies the $M_L=-2$ and $-1$ states, with a nearly equal spin balance such that the total spin moment $M_S^{(z)}(\mathbf{r})$ is close to zero at all $\mathbf{r}$.
This result shows that the smallest energy cost associated with a localization transition in SmN does not preserve Hund's rule.
If the $f$-electrons do undergo a strong rearrangement process in order to satisfy Hund's rule, we expect that this arrangement is associated with an on-site anisotropy cost, as well as a spectroscopic signature, both of which could be observed in experiments.

To a first approximation, treating the magnetic character of the two rigid electrons as a mean ferromagnetic field, the quasiparticle associated with localization would arise from hybridization with the three flexible SOC-like electrons as a three-state system with an effective $J=1$.
In our models, the small spin density of the first empty hole-like natural orbital (\#6) is consistent with a charge-like hybridization process that changes the occupation of the $f$ shell while minimizing the rearrangement of the three existing orbitals.
If this is the case in SmN, it could provide a platform for magnetic pairing of the electrons through fluctuations of a related triplet order parameter with effective $J=1$, a building block for unconventional triplet superconductivity in a ferromagnetic system.

\section{Author contributions}

D.W.T. conceived the project and carried out the main Quanty calculations.
Ø.S.F. and B.G. prepared samples and carried out magnetization measurements.
D.W.T., Ø.S.F., and M.M. designed the Panther experiments and D.W.T. carried them out.
E.X.M.T. carried out DFT calculations and additional Quanty calculations.
D.W.T. wrote the paper with input from all authors.

\section{Data availability}

The ILL data is available at https://doi.ill.fr/10.5291/ILL-DATA.TEST-3456.

\section{Acknowledgements}

We express gratitude to Simon Granville, H. Joe Trodahl, Jackson Miller, and Benedicte S. Ofstad for useful discussions.

\section{Funding}

Ø.S.F. acknowledges funding from the Research Council of Norway (NFR) through project No. 355119.

\section{Methods}

\paragraph{Sample preparation.}

SmN powders were synthesized from dendritic Sm metal (4N) under direct reaction in N2 (5N) flow.
The N2 was purified with an O2 getter.
The reaction was performed at 800 C with intermediate grinding until a homogeneous powder was obtained.

\paragraph{Magnetization measurements.}

Our magnetization measurements were carried out with 64.6~mg of SmN powder loaded into a capsule.
The experiments were carried out in a Quantum Design PPMS in a field of 100~Oe.
The calculated susceptibility was shifted upward by 0.4 units ($0.4 \times 10^{-3}$~emu/mol Sm) to match the baseline of the data, which is most likely caused by the diamagnetic signal from the large capsule.

\paragraph{Panther experiments.}

Inelastic neutron scattering experiments at Panther (Institut Laue-Langevin, France) were carried out with powder loaded into a standard flat plate aluminum can, with a sample area 3 cm tall by 2 cm wide, and approximately 1 mm thickness.
The powder was loaded in a glove bag under inert He gas atmosphere and the can was sealed with indium wire.
The distribution of powder in the can, while mounted vertically, was examined using a neutron camera using the neutron beam at the instrument OrientExpress.
At Panther, a cadmium mask was placed over the leading edge of the sample (facing the incident beam) and over the upper and lower portions of the aluminum holder.
The sample was oriented at a fixed angle in grazing incidence to the neutron beam, which provides an absorption shadow similar to the unconventional $Q$ dependence of the \smthree ion, which reduces our sensitivity to the wavevector dependence but allows us to maintain optimal sensitivity near the peak close to 5~\AA${}^{-1}$.
The ILL experimental data is available at \cite{panther}.

%\clearpage

\end{document}